 \documentclass[nohyper,12pt,letterpaper]{JHEP3}
 \usepackage{epsfig}
 
 \def\bZ{{\mathbb Z}}

 \title{Dyonic D3-branes}
 \author{Bartomeu\,Fiol\\
 Departament de F{\'\i}sica Fonamental i \\Institut de Ci{\`e}ncies del Cosmos, 

Universitat de Barcelona,

Mart{\'\i}\ i Franqu{\`e}s 1, 08193 Barcelona, Catalonia, Spain.\\

\email{bfiol@ub.edu}}

\abstract{We describe embeddings of a single D3-brane in generic IIB supergravity backgrounds when both electric and magnetic fields are turned on. As a particular application, we describe a dyonic D3-brane in $AdS_5 \times S^1$, dual to $1/4$ BPS states of ${\cal N}=4$ SYM.}

\begin{document}

\section{Introduction}
Besides the familiar $1/2$ BPS states, the spectrum of ${\cal N}=4$ super Yang Mills contains $1/4$ BPS states, that in some aspects have much more richer dynamics, since the balance of forces is more delicate, and they can disappear from the spectrum as one varies the vevs of the scalar fields. These states can be labelled by their charges under the Abelian sector of the gauge group, and at least in some regime, can be pictured as consisting of (at least) two $1/2$ BPS cores, separated a certain distance. To construct such configurations, one needs to turn on vevs for at least two scalar fields. Much of what has been learned about these states comes from their realizations as $(p,q)$-string junctions, joining at least three D3-branes in IIB string theory \cite{Bergman:1997yw, Hashimoto:1998zs, Kawano:1998bp, Lee:1998nv}.

In the context of AdS/CFT context, a possible way to represent open Wilson lines associated to $1/2$ BPS particles of ${\cal N}=4$ SYM is through particular D-branes embedded in $AdS_5\times S^5$. This was done for purely electric states in \cite{Rey:1998ik, Drukker:2005kx}, who found a probe D3 brane embedded in $AdS_5$ giving the D3 description of an straight open Wilson line for an electrically charged $1/2$ BPS state; one can readily extend that description to other $1/2$ BPS states in the same $SL(2,\bZ)$ orbit (see also \cite{Gomis:2006sb} for a general realization of $1/2$ BPS Wilson lines in arbitrary representations, in terms of D5 or D3 branes). In this note we generalize that solution, finding D3-brane solutions in $AdS_5 \times S^5$ dual to $1/4$ BPS states of ${\cal N}=4$ SYM. Not surprisingly, since the field configurations have two scalars turned on, the new solutions are embedded in $AdS_5\times S^1$, rather than just $AdS_5$, as it was the case for $1/2$ BPS states.

In fact, it is quite straightforward to describe these embeddings for a much broader type of background: we consider a generic ansatz for a type IIB supergravity solution, of which we are concerned only about a six dimensional part of the metric, and the 5-form flux (we take the rest of the fields to vanish). We show that the equations of motion for a dyonic D3 brane reduce to just the Laplace equation for a three dimensional space. In this way, our ansatz applies to many other backgrounds beyond flat space or $AdS_5 \times S^5$. Our results agree with and generalize those of \cite{Gauntlett:1999xz}, where solutions of the D3 brane were found in the background of stacks of other D3 branes.

Finally we argue that in $AdS_5 \times S^1$ the vev of these Wilson lines is $1$, as it is the case for other Wilson loops preserving $1/2$ or $1/4$ of the supersymmetry \cite{Zarembo:2002an, Guralnik:2004yc, Dymarsky:2006ve}.

\section{Dyonic D3-branes in generic backgrounds}
We consider a IIB SUGRA 10d background  with only metric and $F_5$ turned on. Such types of backgrounds have been studied in detail \cite{Gran:2007ps}. We are going to take a particular ansatz, that as we will later discuss, arises in many IIB supergravity backgrounds constructed from purely D3-branes. Actually, we focus only on a 6d part of the form
\begin{equation}
\begin{array}{l}
ds^2=H^{-1/2}(-dt^2+g_{ij}dx^idx^j)+H^{1/2}(du^2+ dv^2)+\dots  \\
C_{(4)}=H^{-1}dt\wedge dx_1\wedge dx_2 \wedge dx_3
\label{ansatz}
\end{array}
\end{equation}
$H$ does not depend on $t,x^i$. It may depend on $u,v$ and perhaps other dimensions of the 10d background; we are not assuming that $H$ is harmonic in some (perhaps transverse) directions, although that will typicallly be the case. We always identify $t, x^i$ with the world-volume coordinates, and our choice of $C_{(4)}$ is such that in the absence of any world-volume excitations, the full action (Born-Infeld plus WZ term) vanishes.  As we are about to see, purely electric or magnetic D3-branes can be found embedded in a 5d part of this metric, while dyonic D3 branes are embedded in the 6d background. The computations that follow are a generalization of
a number of works \cite{Gibbons:1997xz, Callan:1997kz, Bak:1998xp, Christiansen:1999xq, Gauntlett:1999xz}, where particular choices for $g_{ij}$ and/or $H$ are made.

\subsection{Electric spike}
We turn on $u(x^i)$ and $F_{0i}$. The Born-Infeld determinant is ($2\pi \alpha'=1$)
\begin{equation}
-|G+F|=H^{-2}|g|+H^{-1}|g|g^{ij}(u_iu_j-E_iE_j)-(\nabla u \wedge E)^i g_{ij} (\nabla u 
\wedge E)^j
\end{equation}
The ansatz we take is
\begin{equation}
u_i = E_i
\end{equation}
With this ansatz, it is easy to check that the equations for motion for $A_\mu$ and $u$ both boil down to a single equation
\begin{equation}
\partial _i\left(\sqrt{|g|}g^{ij}u_j\right)=0
\end{equation}
This is the Laplace equation for the 3d manifold with metric $g_{ij}$. Solutions are multicenter harmonic functions in the 3d $x^i$ space.
\subsection{Magnetic spike}
Now we turn on only $v(x^i)$ and the magnetic components of the electromagnetic tensor. The Born-Infeld determinant is now
\begin{equation}
-|G+F|=H^{-2}|g|+H^{-1}(|g|g^{ij}v_iv_j+g_{ij}B^iB^j)+(v_iB^i)^2
\end{equation}
The ansatz we take is
\begin{equation}
v_i=\frac{g_{ij}B^j}{\sqrt{|g|}}
\end{equation}
One can check that with this ansatz the equation of motion for $A_0$ boils down to
$$
\nabla  \wedge (\nabla v) =0
$$
so it is satisfied trivially, while the equation of motion for $v$ is the same as in the electric case
\begin{equation}
\partial _i\left(\sqrt{|g|}g^{ij}v_j\right)=0
\end{equation}

\subsection{Dyonic spike}
Encouraged by the simplicity of these ansatze, we want to study now $1/4$ BPS dyonic case, when both electric and magnetic fields are turned on the world-volume of the D3-brane. As already pointed out in
\cite{Bak:1998xp, Christiansen:1999xq, Gauntlett:1999xz} for particular backgrounds, now one needs to turn on two scalar fields, so we allow for $u(x^i)$ and $v(x^i)$. The Born-Infeld determinant is now
\begin{equation}
\begin{array}{l}
-|G+F|=H^{-2}|g|+H^{-1}|g|g^{ij}(u_iu_j+v_iv_j-E_iE_j)+H^{-1}g_{ij}B^iB^j \\
+(\nabla u\wedge \nabla v)^ig_{ij}(\nabla u\wedge \nabla v)^j -(\nabla u\wedge E)^ig_{ij}(\nabla u\wedge E)^j -(\nabla v\wedge E)^ig_{ij}(\nabla v\wedge E)^j+ \\
H\left(E\cdot (\nabla u\wedge \nabla v)\right)^2 +(u_iB^i)^2+(v_iB^i)^2 -(E\cdot B)^2
\end{array}
\end{equation}
The ansatz is the combination of the previous ones
\begin{equation}
u_i=E_i\hspace{1cm}v_i=\frac{g_{ij}B^j}{\sqrt{|g|}}
\end{equation}
A tedious computation shows that the equations of motion remain as in the previous case
$$
\partial _i\left(\sqrt{|g|}g^{ij}u_j\right)=0\hspace{1cm}
\partial _i\left(\sqrt{|g|}g^{ij}v_j\right)=0
$$

\section{Evaluating the action}
The DBI+WZ lagrangian can be explicitly evaluated for these ansatze. Since the electric and magnetic ones are special cases of the dyonic solution, we just give the result for this case
\begin{equation}
{\cal L}_{DBI+WZ}=\frac{g_{ij}B^iB^j}{\sqrt{|g|}}
\label{lagden}
\end{equation}
We see that the bulk action depends only on the magnetic field, and it vanishes for a purely electric solution. This does not conflict with electric-magnetic duality: the equations of motion are EM dual, while the classical bulk action is not. As it is the case of the evaluation of the action for electrically or magnetically charged black holes,  a boundary term added to the action can restore the duality \cite{Hawking:1995ap}.

\subsection{Boundary terms}
Apart from not being invariant under duality rotations, if we integrate the Lagrangian density (\ref{lagden}) over space, the result will be typically divergent (since everything is independent of $t$, the integral over $t$ also yields a trivial divergence). Both problems are taken of care by recalling that this bulk action has to be supplied with boundary terms \cite{ Drukker:1999zq, Drukker:2005kx}. 
For this ansatz one has that
$$
\frac{\partial {\cal L}}{\partial u_i}=\sqrt{g}g^{ij}u_j \hspace{.5cm}
\frac{\partial {\cal L}}{\partial v_i}=\sqrt{g}g^{ij}v_j \hspace{.5cm}
\frac{\partial {\cal L}}{\partial E_i}=-\sqrt{g}g^{ij}E_j
$$
so using (\ref{lagden}) it follows that
\begin{equation}
{\cal L}-u_i\frac{\partial {\cal L}}{\partial u_i}-v_i\frac{\partial {\cal L}}{\partial v_i}-E_i\frac{\partial {\cal L}}{\partial E_i}=0
\label{legendre}
\end{equation}
{\it i.e.}, if we Legendre transform with respect to the three variables $u,v,A_0$, the full answer is always zero.

\section{Examples}
\subsection{10 dimensional flat space}
Strictly speaking, this is not an example of the backgrounds we wrote down, since there is no RR flux, and no WZ term in the action of the D3 brane. However, since the volume element is constant the only difference is a constant in the action, and the equations of motion we derived also apply in this case. On the other hand, the evaluation of the action is affected by this lack of constant WZ term, and the final result is infinite. 

The case of a single D3 brane in 10d flat space was first considered in \cite {Gibbons:1997xz, Callan:1997kz}, who discussed purely electric or magnetic D3 branes, turning on a single scalar field $X_9$, and generalized in \cite{Bak:1998xp, Christiansen:1999xq}, who turned on two scalar fields  $X^{8,9} (\xi^i)$, and an arbitrary $F_{ab}$, to describe $1/4$ BPS dyons. The BPS bound can be written 
$$
E+iB=e^{i\alpha}\nabla (X^8+iX^9)
$$
where $\alpha$ is a constant arbitary phase. Setting $\alpha=0$ and identifying $X^{8,9}$ with $u,v$ we see this is a particular case of our ansatz. The equations of motion are simply $\nabla ^2 X^8=\nabla ^2 X^9=0$ so the most general solution for each is a multicenter harmonic function. The two center solution
\begin{equation}
X=\frac{q_1}{|\vec r-\vec r_1|}+\frac{q_2}{|\vec r-\vec r_2|} \hspace{.5cm}
Y=\frac{g_1}{|\vec r-\vec r_1|}+\frac{g_2}{|\vec r-\vec r_2|} 
\end{equation}
was argued in \cite{Gauntlett:1999xz} to correspond to a string junction in flat space.

\subsection{$AdS_5 \times S^1$ space}
To find similar solutions in $AdS_5$, one could start by putting a probe D3 in the background of two parallel stacks of D3s, such that the probe has one spike towards each stack  \cite{Gauntlett:1999xz}, and then take the near horizon of one of those stacks. It is perhaps easier to find them directly in $AdS_5 \times S^1$, as a particular case of our general ansatz. Consider $AdS_5 \times S^1$ in  Poincar\'e coordinates. 
\begin{equation}
ds^2=\frac{L^2}{y^2}\; (dy^2-dt^2+d\vec x^ 2)+L^2d\theta ^2
\end{equation}
In this coordinates, the fields we turn on are $y(\vec x)$ and $\theta(\vec x)$. By the change of coordinates $u+iv=\frac{e^{i\theta}}{y}$, we bring this metric to the form (\ref{ansatz}) .
The ansatz for the embedding can be written as
\begin{equation}
\frac{2\pi \alpha'}{L^2}(\vec E+i\vec B)=\vec \nabla (U+iV)
\end{equation}
which saturates the 1/4 BPS bound \cite{Gauntlett:1999xz}. The equations of motion for the scalar fields are simply
\begin{equation}
\nabla ^2 U=\nabla ^2 V=0    
\end{equation}
which has as generic solution
\begin{equation}
U(\vec x)+iV(\vec x)=\frac{e^{i\theta}(\vec x)}{Y(\vec x)}
=\frac{\sqrt{\lambda}}{4N}\sum_{j=1}^n\frac{q_j+ig_j}{|\vec x-\vec x_j|}
\end{equation}
As a first check, if we take $n=1$, then $\theta(\vec x)$ is constant, and the solution is $1/2$ BPS.
Furthermore, for arbitrary $n$, setting all $g_j=0$  or all $q_j=0$, we recover the D3 brane description of respectively a set of $n$ electric or magnetic $1/2$ BPS parallel Wilson lines, which can be embedded enterely in $AdS_5$, since for those particular cases, $\theta(\vec x)$ is constant. The same happens as long as all the charge vectors $(q_j, g_j)$ are parallel among themselves. On the other hand, if two charge vectors $(q_i,g_i)$ and $(q_j, g_j)$ are non-parallel, the solution has a non-trivial $\theta(\vec x)$, corresponding to a $\frac{1}{4}$ BPS dyon, and it's embedded in $AdS_5 \times S^1$.

The bulk contribution to the action, given by the integral of (\ref{lagden}), is divergent for $B\neq 0$. For a purely electric spike, the bulk contribution is zero. It was argued in \cite{Drukker:2005kx} that in this case one has to Legendre transform with respect to $u$ and $A_0$, and these two boundary terms cancel each other, as it follows also from  (\ref{legendre}) since in this case $v$ is not turned on. For a purely magnetic spike, the bulk term is not zero, but it gets cancelled by the Legendre transform with respect to $v$. Finally for a a dyonic solution, the bulk term is non-zero but after considering the Legendre transform with respect to $u,v$ and $A_0$ the full action is zero.

\subsection{Other backgrounds}
The ansatz we took for a six dimensional piece of a ten dimensional IIB SUGRA background (\ref{ansatz}) is realized by many other solutions, beyond flat space and $AdS_5 \times S^5$, {\it e.g.} the backgrounds that desribe the Coulomb branch of ${\cal N}=4$ SYM \cite{Freedman:1999gk} or even backgrounds where the $3d$ metric $g_{ij}$ is no longer $\delta_{ij}$, {\it e.g.} those corresponding to intersecting D3-branes  \cite{Tseytlin:1996bh}. One issue that might be worth studying is whether one can reproduce the lines of marginal stability for the $1/4$ BPS states in the Coulomb branch of ${\cal N}=4$, by considering these dyonic D3 branes in the backgrounds of \cite{Freedman:1999gk}.

\section{Acknowledgements} 
I would like to thank Nadav Drukker for comments and discussions. This work was supported in part by by grants FPA2007-66665C02-02 and  DURSI 2005-SGR-00082 and by a Ram{\'o}n y Cajal fellowship.

\end{document}